\documentclass[12pt]{article}

\usepackage[T1]{fontenc}
\usepackage{graphicx}
\usepackage{latexsym}
\usepackage{amsmath,a4wide, amssymb, amsfonts, graphics, epsfig, makeidx, subfigure}
\usepackage{mathrsfs}
\usepackage{comment}

\usepackage{hyperref}

\usepackage{caption}
\usepackage{changepage}
\usepackage{booktabs}

\newlength{\extralength}
\newcommand{\acknowledgments}[1]{
\vspace{6pt}\noindent{\fontsize{9}{11.2}\selectfont\textbf{Acknowledgments:} {#1}\par}}

\def\@reftitle{}
\newcommand{\reftitle}[1]{\gdef\@reftitle{#1}}

\makeatletter
\@addtoreset{equation}{section}
\makeatother

\newcommand{\ZZ}{\mathbb Z}
\newcommand{\RR}{\mathbb R}
\newcommand{\CC}{\mathbb C}
\newcommand{\EE}{\mathbb E}			
\newcommand{\PP}{\mathbb P}			
\newcommand{\IF}[1]{\mathbf{1}_{#1}}
\newcommand{\eq}{ \!=\!}
\newcommand{\plus}{\!+\! }
\newcommand{\minus}{\!-\! }
\newcommand{\mdot}{\!\cdot\! }
\newcommand{\cc}{\; | \;}
\newcommand{\Cc}{\; \big| \;}

\newcommand{\rd}{{\rm d}}
\newcommand{\ri}{{\rm i}}

\newcommand{\dg}[1]{\boldsymbol{#1}}
\newcommand{\lh}{\dg{\big{\langle}}} 
\newcommand{\rh}{\dg{\big{\rangle}}}
\newcommand{\tens}{\otimes}

\newcommand{\bfe}{{\bf{e}}}
\newcommand{\bff}{{\bf{f}}}

\title{A Probability Model  for the Bell Experiment
}
\author{ Kees van Hee \and Kees van Berkel \and Jan de Graaf \footnote{ Jan de Graaf passed away 2024, September 16}}

\date{
    {\bf Also appeared in Quantum Reports, Volume 8, Issue 1, 2026} 			\\ \vspace{5mm}
    Technical University Eindhoven, Dept. of Mathematics \& Computer Science	\\
    P.O. Box 513, 5600 MB Eindhoven, The Netherlands  					\\
    email:  \{k.m.v.hee, c.h.v.berkel\} @tue.nl           						\\ \vspace{5mm}
    \today
}

\begin{document}

\maketitle

\abstract{
The Bell inequality constrains the outcomes of measurements on pairs of distant entangled particles.
The Bell contradiction states that the Bell inequality is inconsistent
with the calculated outcomes of these quantum experiments.
This contradiction led many to question the underlying assumptions, viz. so-called realism and locality.
The probability model 
underlying the Bell inequality is generally left implicit.
{We}  
propose an explicit probability model for the CHSH version of the Bell experiment.
This model has only two simultaneously observable detector settings per measurement,
and therefore does not assume realism.
The quantum expectation now becomes a \emph{{conditional}} expectation, \emph{{given}}  the two detector settings.
This probability model is in full agreement with both quantum mechanics and experiments.
As a result, the model satisfies the Bell inequality; there are no so-called violations.
We extend this model to include a hidden variable.
This extended model is not Bell-separable.
This non-separability implies that the model is non-deterministic or non-local (or both).}

{\bf Keywords:} Bell inequality, CHSH inequality, Bell experiment, Bell contradiction, EPR-paradox, 
separability, hidden variables.

\section{Introduction}
\label{sec:introduction}

J.S. Bell  describes an experiment~\cite{Bell:1964} in which ``a pair of spin one-half particles,
formed somehow in a singlet spin state and moving freely in opposite directions''.
This experiment was inspired by the EPR paradox~\cite{Einstein:1935}.
He derives an inequality, labeled (15) in~\cite{Bell:1964},
that constrains the expected outcomes of the measurements of selected components of the spins of these two particles.
Today this inequality is known as the {\em {Bell inequality}}.

The Bell inequality connects two different formalisms: quantum mechanics and probability models.
In quantum mechanics the state of a system is modeled as a vector in a Hilbert space.
The evolution of this state is governed by a unitary operator and is deterministic.
An observation (measurement) of this state is by the application of a Hermitean operator which yields
a probability value for each eigenvector of that operator.
This measurement procedure is inherently probabilistic (non-deterministic)
and thus requires a (formal) probability~model.

In the body of this paper we focus on the more general CHSH version~\cite{Clauser:1969} of the Bell experiment
with four detector settings, two detector on each site.
The spins of the two prepared particles (photons) are measured
by detectors (polarimeters) {$A$}  and {$B$.}  {(The main findings of our work can straightforwardly be adapted 	to the original Bell experiment with three detector settings and to fermions.)}  
Each detector has two different settings,
and the selection among the available detector settings is possibly random,
and possibly {\em {after}}  the photon pair has been prepared.
Each individual measurement outcome has either value +1 (spin up) or {$-$}1  
(spin down).
The corresponding stochastic variables for the detector settings of detector $A$
are denoted by $X_0$, $X_1$, and~for detector $B$ by $Y_0$, and~$Y_1$.
The CHSH version~\cite{Clauser:1969} of the Bell inequality is:
\begin{equation}
\Big| \EE [X_0 Y_0] +  \EE[X_1 Y_0] +  \EE[X_1 Y_1] -  \EE[X_0 Y_1] \Big| \leq 2 ~,
\label{eq:CHSH-intro}
\end{equation}
where $\EE[Z]$ denotes the expectation of random variable $Z$, see Section~\ref{basics}.

Next, Bell also calculates the outcomes of these experiments,
by applying the relevant observables (self-adjoint operators) to the quantum state of the photon pair.
He thus arrives at the Bell contradiction: these calculated outcomes contradict the inequality.
The Bell inequality also inspired numerous experiments, including~\cite{Aspect:1982,Hensen:2015}.
These experiments typically demonstrate ``loophole-free Bell-inequality~violation''.

The Bell contradiction led many to question the two assumptions made in the proof for
Inequality (\ref{eq:CHSH-intro}), see~\cite{Nielsen:2000}:
\begin{enumerate}
\item The underlying physical properties for observations $X_0$, $X_1$, $Y_0$, $Y_1$
exist independently of being observed or measured.
This assumption is known as {\em {realism}}   or as {\em {counterfactual definiteness}}.
\item The setting of detector A does not impact the outcome Y on the other detector
and vice~versa.
This assumption is known as  {\em {(statistical) locality}},  see Section \ref{subsec:statistical_locality}.
\end{enumerate}

{The}  
details of these assumptions are still being widely debated
among physicists, probability theorists, and~philosophers,
see for example~\cite{Gill:2014, Brunner:2014, Hall:2015}.
{To date,} Ref.~{\cite{Bell:1964}}  {has been cited over 16 thousand times.}  

In this paper we offer a different and fresh perspective on the Bell contradiction.
Rather than questioning the underlying assumptions, we question the underlying probability model itself.
A probability model comprises an {\em {outcome space}},  an~{\em {event space}}  and a {\em {probability measure}}.
These notions will be reviewed in some detail in Section~\ref{basics}.
Interestingly and significantly, a~suitable outcome space for the Bell experiment is generally left implicit.
The few exceptions include~\cite{Hensen:2015, Chaves:2017, Gill:2021},   
where the experiment is modeled as a Bayesian network.
However, these Bayesian-network models are inherently factorizable with conditional probabilities,
and therefore disagree with quantum mechanics.
In Section~\ref{sec:quantum}, we derive the quantum mechanical description of the Bell experiment.
In Section~\ref{wrong model} we review the model with four observables
that seems to have been used implicitly by other researchers and that leads to the Bell paradox.
Bell~\cite{Bell:1964} derived his inequality in a similar way.
Next, in~Section~\ref{suitable model}, we present a suitable probability space for the Bell experiment,
including a probability measure, derived by quantum mechanics in the former section,
and calculate the expected values of the observables in the Bell inequality.
In this model there is no Bell~contradiction.

Bell introduced a so-called {\em {hidden variable}}
to see if it could account for the observed correlations among measurement outcomes.
In Section~\ref{sec:model-with-hidden-variables} we extend the proposed model to include
such a hidden variable.
The probability measure of this extended model is shown non-separable in the Bell-sense.
Furthermore, this non-separability implies that the model is non-local, or~non-deterministic, or~both.
Appendix \ref{alternative} offers an alternative proof of this~non-separability.

\section{The Basics of Probability~Theory}
\label{basics}
The basics of probability theory can be summarized as~follows.
\begin{itemize}

\item  A {\em {probability space}}  \cite{Kolmogorov:1933,feller:1968,wiki:probability-space}
is a 3-tuple  $(\Omega, \cal F, \PP )$.
Here $\Omega$ is an arbitrary set called the {\em {outcome space}}  (or sample space),
$\cal F$ is the {\em {event space}}  (formally a $\sigma$-algebra).
The last component $\PP $ is a {\em {probability measure}}  (or probability) on $\cal F$.

\item A {\em {stochastic variable}}  is defined as a (measurable) function on a probability space \linebreak
$X:\Omega\rightarrow \Sigma$, where $\Sigma=\RR$ or $\Sigma$ is countable.
The probability $\PP[\{\omega \in \Omega\} \cc X(\omega) \eq x]$ is abbreviated as $\PP[X \eq x]$
and called the {\em {distribution}}  of $X$.

\item Given event $A \in \cal F$, the~{\em {conditional probability}}  of $X$ given $A$ is denoted by $\PP[X | A]$
and is defined by $\PP[X | A] := \PP[X \cap A]/\PP[A]$.
This notion can be generalized for conditioning on a sub-$\sigma$-algebra of $\cal F$.

\item Two events $A$ and $B$ are called {\em {independent}}
if $\PP[A\cap B]=\PP[A]\cdot \PP[B]$ and similarly two stochastic variables $X$ and $Y$
are {\em {independent}}  if $\PP[X\in A , Y \in B]=\PP[X \in A] \cdot\PP[Y \in B]$ for all events $A,B \in\cal F$.

\item The {\em {expectation}}  of a random variable $X$ is $\EE[ X]= \sum_{x} x.\PP[X=x]$
and for continuous random variables $\EE[X]= \int_{x}x.p(x)dx$ where $p(x)=\frac {d}{d x}\PP[X\leq x]$
is the probability {\em {density}}.  And~similarly for conditional expectations:
$\EE [X\cc A]=\sum_{x} x.\PP[X=x\cc A]$.

\end{itemize}

{The}  
only relationship between probability theory and experimental observations
is given by the law of large numbers  and the central limit theorem.
The first one states that the  average  of $n$ independent and identical distributed,
real-valued stochastic variables converges to a theoretical value, the expectation.
The second one states that the distribution of this average
(normalized by subtracting the average and divided by the variance of the average)
converges to the Gauss distribution.
Both laws have formal proofs and are empirically validated with high~precision.

\section{A Quantum Mechanical~Model}
\label{sec:quantum}

\begin{figure}[!ht]
\begin{center}
\includegraphics[width=0.8\textwidth]{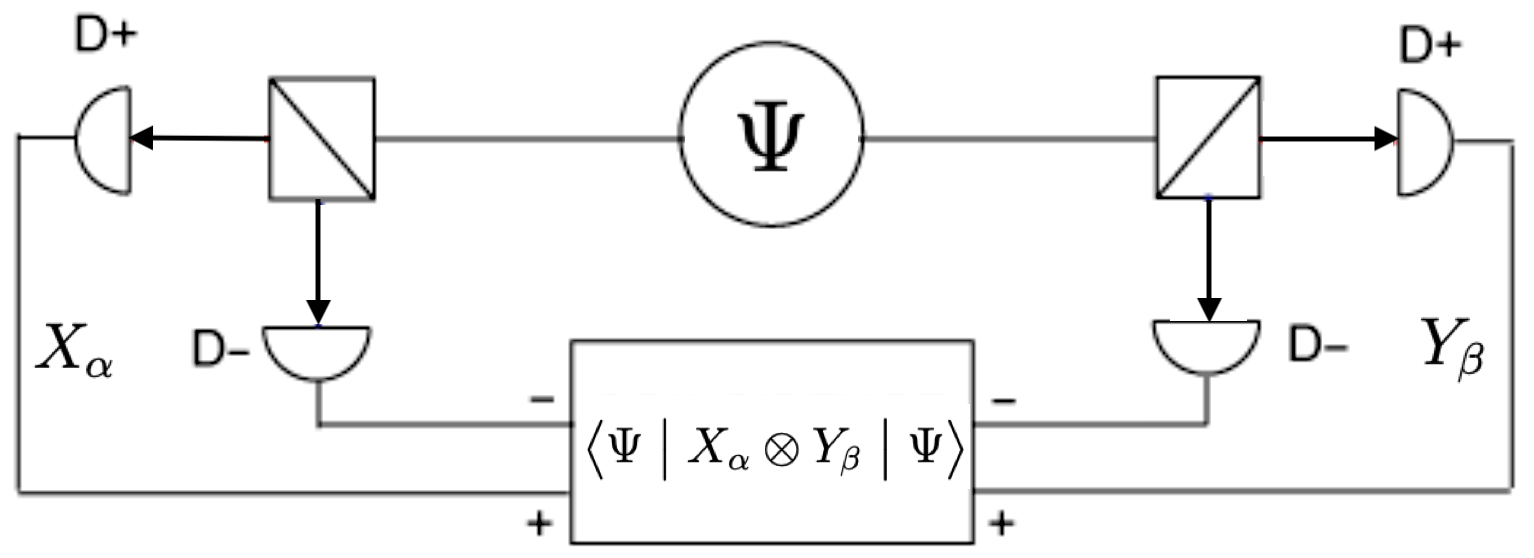}
\end{center}
\vspace{-5mm}
\caption{The {optical}  version of the Bell experiment.
The entangled photon pair is described by wave function $\Psi$.
The polarimeters $X_\alpha$ and $Y_\beta$ each comprise a polarizer,
with rotatable polarization angles $\alpha$ and $\beta$, and~two photon counters labeled D+ and D{$-$.}
The coincidence monitor $\lh \Psi\Cc X_\alpha \tens Y_\beta \Cc \Psi \rh$
computes the corresponding quantum mechanical expectation.
}
\label{fig:bell}
\end{figure}

In a quantum mechanical model of the Bell experiment, depicted in Figure~\ref{fig:bell},
there are two detectors,  polarimeters, in~a line, that we call the y-axis. The~detectors measure the spin in the x-z-plane perpendicular to the y-axis. There are two base vectors $\bfe_0=(1,0)^\top$ and $\bfe_1=(0,1)^\top$ used for the spin also known as up and down.  The~quantum state of an entangled photon pair can be described by
\begin{equation}
\dg{\Psi} = \frac1{\sqrt2}(\bfe_0 \tens \bfe_1-\bfe_1 \tens \bfe_0)   ~\in~ \CC^2 \tens \CC^2~.
\end{equation}
{The}  
polarization  is specified by a vector $\bff_\theta=(\cos(\theta),\sin(\theta))^\top$ which makes an angle $\theta$ with the x-axis. We define  $\tilde{\bff}_\theta := \bff_{\theta+\frac\pi2}$, so
$\tilde{\bff_{\theta}}=(-\sin(\theta),\cos(\theta))^\top$, which is orthogonal to $\bff_\theta$.
The measurement of a photon is done by an observable  which could be the projection operator:
$\bff_\theta.\bff_\theta^\top$. Note that $\bff_\theta.\bff_\theta^\top.\bff_\theta=\bff_\theta$.
So, the~eigenvalue is $1$ and $\bff_\theta$ is the eigenvector of this projection operator.
The only other eigenvalue of this operator is $0$ with eigenvector $0$. We like to have two non-zero eigenvalues as measurements for the spin up and down. Therefore we transform the projection operator into $2\bff_\theta.\bff_\theta^\top-\bf{I}$ where $\bf{I}$ is the identity operator on the x-z-plane. Now the eigenvalues are $1$ and $-1$ with eigenvectors $\bff_\theta$ and $\tilde{\bff_\theta}$. It is easy to verify that
\begin{equation}
F_\theta:=\bff_\theta.\bff_\theta^\top-\tilde\bff_\theta.\tilde{\bff_\theta}^\top=2\bff_\theta.\bff_\theta^\top-\bf{I}
\end{equation}
{We}  see that $F_{\! \theta} :\CC^2 \to \CC^2$, with~$\theta \in [0,\pi)$ is a special type of self-adjoint operator (Hermitean matrix), which is the difference of two orthogonal projections. So, we derived
\begin{equation}
F_{\theta}    		= \begin{bmatrix} 	\cos(2\theta) & \sin(2\theta) \\
\sin(2\theta) & -\cos(2\theta) 	\end{bmatrix},~~~
\bff_\theta    		= \begin{bmatrix} \cos \theta \\ \sin \theta  						\end{bmatrix},~~
\tilde{\bff}_\theta 	= \bff_{\theta+\frac\pi2}
= \begin{bmatrix} -\sin \theta \\~\cos \theta  						\end{bmatrix}~.
\end{equation}
with eigenvalues (measurements)  are  $+1$ and $-1$, with~respective eigenvectors $\bff_\theta$ and $\tilde{\bff}_\theta$.
The CHSH experiment has two detectors, one left $F_\alpha$ and one right $F_\beta$ and together they form one operator, the~observable,  modeled
by $F_{\alpha} \tens F_{\beta}~$ acting on the entangled state  $\dg{\Psi}$.

In a quantum mechanical model of the Bell experiment
the quantum state of an entangled photon pair can be described by
\begin{equation}
\dg{\Psi} = \frac1{\sqrt2}(\bfe_0 \tens \bfe_1-\bfe_1 \tens \bfe_0)   ~\in~ \CC^2 \tens \CC^2~.
\end{equation}
{Here}  $\bfe_0=(1,0)^\top$ and $\bfe_1=(0,1)^\top$ the two base spin vectors, also known as up and down.

The spin of a photon can be detected by a polarimeter,
which is modeled by a special type of self-adjoint operator (Hermitean matrix)
$F_{\! \theta} :\CC^2 \to \CC^2$,
\begin{equation}
F_{\theta}    		= \begin{bmatrix} 	\cos(2\theta) & \sin(2\theta) \\
\sin(2\theta) & -\cos(2\theta) 	\end{bmatrix} ~,
\end{equation}
where $\theta \in [0,\pi)$ is the angle of the polarizer with respect to some reference frame perpendicular to the axis of the experiment.
$F_{\theta}$ is the difference of two orthogonal projections
\begin{equation}
F_{\theta}			= \bff_\theta \bff_\theta^\top-\tilde{\bff}_\theta\tilde{\bff}_\theta^\top\!,~~~
\bff_\theta     		= \begin{bmatrix} \cos \theta \\ \sin \theta  						\end{bmatrix},~~
\tilde{\bff}_\theta 	= \bff_{\theta+\frac\pi2}
= \begin{bmatrix} -\sin \theta \\~\cos \theta  					\end{bmatrix}~.
\end{equation}
{The}  outcomes (eigenvalues) of a detection are  $+1$ and $-1$,
with respective eigenvectors $\bff_\theta$ and $\tilde{\bff}_\theta$.
Note that this covers all cases, because~$\bff_{\theta\,\pm\,\pi}=-\bff_\theta$.

The CHSH experiment has two detectors
$F_\alpha$ on the left hand side  and $F_\beta$ on the right hand side.
For later convenience, these are renamed to $X_\alpha$ and $Y_\beta$ respectively.
The outcome of the Bell experiment  can now be modeled
by $X_{\alpha} \tens Y_{\beta}~$ acting on the entangled state  $\dg{\Psi}$.
For the probability calculations, it is advantageous to orthogonally decompose
the entangled state $\dg{\Psi}$  according  to the spectral resolution of $X_{\alpha} \tens Y_{\beta}~$,
\begin{equation}
\dg{\Psi}	= 	\psi_{1,1}\,	   \bff_\alpha	\tens		\bff_\beta
\;+\; 	\psi_{-1,1}\,\tilde{\bff}_\alpha 	\tens		\bff_\beta
\;+\; 	\psi_{1,-1}\,	    \bff_\alpha	\tens \tilde{\bff}_\beta
\;+\;	\psi_{-1,-1}\,\tilde{\bff}_\alpha	\tens	\tilde{\bff}_\beta ~.
\end{equation}
{The}  coefficients $\psi_{x y},~x,y \in \{-1,1\}$ are calculated by taking inner products (in Dirac notation). A~specimen,
\begin{equation}
\psi_{-1,1}		= \lh\dg{\Psi} \dg{,} \; \tilde{\bff}_\alpha \tens \bff_\beta\rh_{\CC^2\tens\CC^2}~.
\end{equation}

The probability of the outcome pair $(x, y)$ for detection angles $\alpha$ and $\beta$ is given by Born's rule
\begin{equation}
p(x, y \Cc \alpha, \beta)  ~=~ |\psi_{xy}|^2 ~.
\label{eq:wave-prob}
\end{equation}
{In}  Section~\ref{suitable model}  we will see that this is indeed a {\em {conditional probability}},  given the detector settings.
A detailed specimen,
\begin{equation}
\begin{array}{lll}
p(-1, 1 \Cc \alpha, \beta)
=	| \psi_{-1,1} |^2
&=	\frac12 \Big(\det \begin{bmatrix} 	\bfe_0^\top \tilde{\bff}_\alpha&\bfe_0^\top \! \bff_\beta	\\[1.5mm]
\bfe_1^\top \tilde{\bff}_\alpha&\bfe_1^\top \! \bff_\beta 	\end{bmatrix} \Big)^2 \\[6mm]
&=	\frac12 \Big(\det \begin{bmatrix}	 	- \sin \alpha 	&\cos \beta		\\
~\cos \alpha 	& \sin \beta  		\end{bmatrix} \Big)^2
=	\frac12 \cos^2(\alpha \minus \beta).
\end{array}
\label{eq:Pxy-a}
\end{equation}
{In}  the same way one obtains the following full results:
\begin{equation}
\begin{array}{lll}
p( 1, 1 \Cc \alpha, \beta)  	&= p(-1, -1 \Cc \alpha, \beta) 	&= \frac12\sin^2(\alpha \minus \beta)  	\\[1mm]
p(-1, 1 \Cc \alpha, \beta)  	&= p( 1, -1 \Cc \alpha, \beta) 	&= \frac12\cos^2(\alpha \minus \beta) ~.
\end{array}
\label{eq:Pxy-b}
\end{equation}
{Given}  these probabilities, the~ quantum mechanical {\em {expectation}}  $\lh X_\alpha \tens Y_\beta \rh$
can now be calculated as follows
\begin{equation}
\begin{array}{lll}
& \lh \Psi\Cc X_\alpha \tens Y_\beta \Cc \Psi \rh 										\\[1mm]
= &  (1 \mdot 1) \mdot p( 1, 1 \Cc \alpha, \beta) ~+~ (-1 \mdot -1)\mdot  p( -1, -1 \Cc \alpha, \beta) + 	\\[1mm]
&  (-1 \mdot 1)\mdot  p(-1, 1 \Cc \alpha, \beta) ~+~ (1 \mdot -1)\mdot  p( 1, -1 \Cc \alpha, \beta) 		\\[1mm]
= & \sin^2(\alpha \minus \beta) - \cos^2(\alpha \minus \beta) 								\\[1mm]
= & 	-\cos\big(2(\alpha-\beta)\big) ~.
\end{array}
\label{eq:expectation}
\end{equation}
{This}  expectation is in agreement with those of~\cite{Clauser:1969,Aspect:1982}.
({{The}  
minus sign preceding the $\cos()$ is sometimes omitted due to different conventions.})
{Figure~3}  
in~\cite{Aspect:1982} shows an excellent agreement between the measured data and (\ref{eq:expectation}).

\section{A Model with Four~Observables}
\label{wrong model}
In nearly all publications on the Bell experiment, including Bell's, the~probability model underlying the experiment is left implicit.
That is, they lack a probability model consistent with the standard definitions given in Section~\ref{basics}.
These implicit models tacitly assume four stochastic variables (observables) $X_0, X_1,Y_0, Y_1$,
corresponding to the four detector settings $a_0, a_1, b_0, b_1$.
During a particular experiment only two of them are observed
(one $X$ on the left-hand side and one $Y$ on the right-hand side),
but the other two are assumed to exist as well.
This assumption is sometimes called {\em {realism}}  \cite{Gill:2021}.
Note that these the values of the four random variables are in $\{-1,1\}$.
The outcome space $\Omega$ for this experiment is given in Table~\ref{tab:candidate-outcome-space}.

\begin{table}[htbp]
  \centering
    \begin{tabular}{r|rrrrrrrrrrrrrrrr}         \hline
        		& 1 	& 2 	& 3 	& 4 	& 5	& 6 	& 7 	&  8  	& 9 	& 10 	& 11	& 12	& 13	& 14	& 15	& 16	\\    \hline
    $X_0$   & -1 	&  1  	& -1 	&  1	& -1	&  1  	& -1 	&  1  	& -1 	&  1  	& -1 	&  1	& -1	&  1 	& -1	& 1	\\   
    $X_1$   & -1 	& -1 	&  1 	&  1 	& -1	& -1	&  1 	&  1 	& -1 	& -1 	&  1 	&  1 	& -1	& -1	&  1	& 1 	\\   
    $Y_0$   & -1 	& -1	& -1 	& -1	&  1 	&  1  	&  1 	&  1  	& -1	& -1	& -1	& -1	&  1	&  1	&  1	& 1	\\   
    $Y_1$   & -1 	& -1 	& -1 	& -1 	& -1 	& -1	& -1	& -1 	&  1 	&  1 	&  1	&  1 	&  1 	&  1	&  1	& 1	\\   \hline                                                             
    CHSH  &  2 	&  2  	& -2 	& -2	& -2 	&   2 	& -2 	&  2 	&  2 	& -2 	&  2 	& -2 	& -2 	& -2 	&  2 	& 2 	\\  \hline                                                               
    \end{tabular}
   \newline
   \caption{An outcome space for the CHSH experiment.}
  \label{tab:candidate-outcome-space}
\end{table}

The event space $\mathscr{F}$ is the power set of $\Omega$.
We leave the probability measure unspecified, because~the arguments we present hold for all possible probabilities measures.
It is based on a simple inequality~\cite{Nielsen:2000}: for $Q,R,S,T\in \{-1,1\}$
\begin{equation}
\big| QS+QT+RS-RT\big|  \leq 2 ~.
\label{eq:CHSH-inequality}
\end{equation}
{This}  inequality can be generalized to $Q,R,S,T\in [-1,1]$. First we note that
\begin{equation}
\big| QS+QT+RS-RT\big|
~\leq~ 	\big|Q(S+T)\big|+\big| R(S-T)\big|
~\leq~ 	 \big|S+T\big|+\big|S-T\big|~.
\end{equation}
{The}  right-hand side equals  $2\big| S\big|$ or $2\big| T\big|$,
depending on the four cases of the signs of $S+T$ and $S-T$.
So, for~these four random variables we have (see last row in Table~\ref{tab:candidate-outcome-space}):
\begin{equation}
\big| X_0 Y_0 + X_1 Y_0 +  X_1 Y_1 - X_0 Y_1\big| ~\leq~ 2 ~.
\label{eq:Bell-inequality1}
\end{equation}
{By}  taking expectations (a linear operation) we obtain
\begin{equation}
\big|\EE [X_0 Y_0] + \EE [X_1 Y_0] + \EE[ X_1 Y_1] -\EE[ X_0 Y_1]\big|  ~\leq~ 2 ~.
\label{eq:Bell-inequality}
\end{equation}
{Note}  that it $\big|\EE[X_\alpha Y_\beta \big|\leq 1$ and so, for~{\em {any}}  probability measure
on the outcome space of Table~\ref{tab:candidate-outcome-space} Equation~(\ref{eq:Bell-inequality}) holds.
Using (\ref{eq:expectation}) it follows that
\begin{equation}
\EE[X_\alpha Y_\beta]=\lh \Psi | X_\alpha\tens Y_\beta |~\Psi\rh=\lh X_\alpha\tens Y_\beta\rh=-\cos(2(\alpha-\beta)) ~.
\label{eq:inequality1}
\end{equation}
{Note}  that we use here the symbols  $X$ and $Y$ for the random variables as well as for the operators $X$ and $Y$ they relate~to.

Next, we will use the specific  values for the detectors $\alpha: a_0 ,a_1$ and two for $\beta: b_0, b_1$.
(Recall that these settings correspond to the random variables $X_0, X_1, Y_0, Y_1$ as used in
Table~\ref{tab:candidate-outcome-space}).
Following~\cite{Nielsen:2000},
\begin{equation}
a_0=0, ~a_1=\!\frac14 \pi, ~b_0=\!\frac58 \pi, ~b_1=\!\frac78 \pi ~.
\label{angles}
\end{equation}
{These}  represent the so-called Tsirelson's bound~\cite{Pal:2022},
where the Bell inequality allegedly is violated most.
From (\ref{eq:inequality1}) it follows that
\begin{equation}
\lh X_0 \tens Y_0 \rh = \lh X_1 \tens Y_0 \rh = \lh X_1 \tens Y_1 \rh = -  \lh X_0 \tens Y_1 \rh =\frac{1}{2}\sqrt{2}~,
\label{eq:CHSH-Born1}
\end{equation}
which results in
\begin{equation}
\big| \lh X_0 \tens Y_0 \rh + \lh X_1 \tens Y_0 \rh + \lh X_1\tens  Y_1 \rh -  \lh X_0 \tens Y_1 \rh \big| = 2 \sqrt{2}~.
\label{eq:CHSH-Born}
\end{equation}
{With}  the {\em {assumption}}
\begin{equation}
\EE[X_iY_j]=\lh X_i\tens Y_j \rh ~\text{for}~i,j \in\{0,1\}
\label{eq:wrong expectation}
\end{equation}
there is a manifest contradiction between (\ref{eq:CHSH-Born}) and (\ref{eq:Bell-inequality}).
Our conclusion from this contradiction is that assumption (\ref{eq:wrong expectation}) does not hold.
More generally, the~probability model with four stochastic variables $X_0,X_1,Y_0,Y_1$
does not fit the quantum mechanical model of the experiment.
There are only two observables and two detector settings.
This means that $\lh X_\alpha \tens Y_\beta \rh$ must be a {\em {conditional}}  expectation,
{\em {given}}  the detector settings, as~will be explained in the next~section.

\section{A Model with 2 Observables and 2 Detector~Settings}
\label{suitable model}

A suitable probability model for the Bell experiment can be based on a different set of four stochastic variables:
the two {\em {observables}}  $X$ and $Y$ (the two measurements) and the two {\em {settings}}   $A$ and $B$.
Furthermore, the~quantum mechanical expectation $\lh X_\alpha \tens Y_\beta \rh$
now becomes the {\em {conditional}}  expectation
\begin{equation}
\lh X_\alpha \tens Y_\beta \rh=\EE\big[XY \cc A=\alpha, B=\beta\big]~,
\end{equation}
that is, the~expectation to observe  $X$ and $Y$ {\em {given}}  detector settings $A$ and $B$.
The corresponding (conditional) probability space is shown in Table~\ref{tab:conditional-probability-measure}.
Note that the outcomes of $A$ and $B$ must be known in order to compute the simultaneous probability of $(X,Y)$.
The tabulated probabilities are based on (\ref{eq:Pxy-b}),
and hence agree with the quantum mechanical model and its experimental validation~\cite{Aspect:1982}.
\begin{table}[htbp]
  \centering
  \renewcommand{\arraystretch}{1.2}         
    \begin{tabular}{p{6mm}p{6mm}|p{25mm}p{10mm}}				\cline{1-3}
           	&             	& $\alpha$		              		& $A$ 	\\
     $X$ 	& $Y$     	& $\beta$ 						& $B$ 	\\       \cline{1-3}  
    1 &  1 	& $ \frac12\sin^2 (\alpha \minus \beta)$  		& 		\\  	 
   -1 &  1 	& $ \frac12\cos^2 (\alpha \minus \beta)$ 		& 		\\ 	 
    1 & -1 	& $ \frac12\cos^2 (\alpha \minus \beta)$ 		&  		\\   	 
   -1 & -1 	& $ \frac12\sin^2 (\alpha \minus \beta)$    	   	&  		\\  	\cline{1-3}
  \end{tabular}
\newline
 \caption{The  conditional probability measure given $A=\alpha, B=\beta$.}
  \label{tab:conditional-probability-measure}
\end{table}

From Table~\ref{tab:conditional-probability-measure} it can be observed that
\begin{equation}
\begin{array}{lllll}
\PP[X \eq x  \;|\;  A \eq \alpha, B \eq \beta] 	&=& \PP[X \eq x  \;|\;  A \eq \alpha] 	&=& \frac12 	\\[1mm]
\PP[Y \eq y  \;|\;  A \eq \alpha, B \eq \beta] 	&=& \PP[Y \eq y  \;|\;  B \eq \beta] 	&=& \frac12 	~.
\end{array}
\label{eq:locality}
\end{equation}
{Clearly,}  the~outcome $X$ does not depend on the setting of detectors $A$ and $B$.
Note that this property holds for all detector settings and for any probability $\PP[A \eq \alpha, B \eq \beta]$.

When we fill in the specific parameters from (\ref{angles}) we obtain Table~\ref{tab:CHSH-probability-measure},
which depicts the complete probability space:  $\Omega = \{-1,1\}^2 \times \{a_0,a_1\} \times \{b_0,b_1\} $
(a set of 16 cells), $\mathscr{F}$ is the power set of it and $\PP[.]$ is represented by the content of the cells.
It is easy to verify that $\PP[A\!=\!a_i,B\!=\!b_j]= \frac{1}{4}$ for $i,j\in \{0,1\}$,
which means that $A$ and $B$ are  {\em {independent}. }

\begin{table}[htbp]
  \centering
     \begin{tabular}{p{6mm}p{6mm}|p{12mm}p{12mm}p{12mm}p{12mm}p{6mm}}							\cline{1-6}
           	&             	& $a_0$			& $a_1$ 			& $a_1$  			& $a_0$   		& $A$ 	\\
     $X$ 	& $Y$     	& $b_0$ 			& $b_0$ 			& $b_1$  			& $b_1$ 		& $B$ 	\\ \cline{1-6}  
    1     	&  1    	& $\gamma^2/8$  	& $\gamma^2/8$   	& $\gamma^2/8$  	& $\beta^2/8$        	&  	\\[0.5mm]  
   -1    	&  1     	& $\beta^2/8$      	& $\beta^2/8$       	& $\beta^2/8$       	& $\gamma^2/8$   	&  	\\[0.5mm]  
    1     	& -1   	& $\beta^2/8$        	& $\beta^2/8$       	& $\beta^2/8$       	& $\gamma^2/8$   	&  	\\[0.5mm]  
   -1    	& -1    	& $\gamma^2/8$   	& $\gamma^2/8$  	& $\gamma^2/8$  	& $\beta^2/8$        	& 	\\[0.5mm] \cline{1-6}
  \end{tabular}
  $\begin{matrix}~~ \\[10mm] 
  	\beta^2		= \sin^2(\frac\pi8)	= \frac14(2-\sqrt2)  	\\[1.5mm] 
   	\gamma^2		= \cos^2(\frac\pi8)	= \frac14(2+\sqrt2)  
   \end{matrix} $
\newline 
\caption{The Bell probability measure for specific detector settings and for all detector combinations the same probability.} 
  \label{tab:CHSH-probability-measure}
\end{table}
Note that the first three columns are identical,
because  $\sin^2(0  \minus  \frac58\pi)  =  \sin^2(\frac14\pi  \minus  \frac58\pi)  =  \sin^2(\frac14\pi  \minus  \frac78\pi) = \cos^2(\frac18\pi) = \gamma^2$.
Similarly, we derive for $(i,j) \in \{(0,0), (1,0),(1,1)\}$:
\begin{equation}
\begin{array}{llrrrr}
\EE[XY \Cc  A\!=\!a_i,B\!=\!b_j]	  =  -\cos (2(a_i-b_j)) 	&=& \cos(\frac{1}{4}\pi) 	&=& \frac{1}{2}\sqrt (2). ~, \\[1mm]
\EE[XY \Cc  A\!=\!a_0,B\!=\!b_1]				&=& -\cos(\frac{1}{4}\pi)  	&=& -\frac{1}{2}\sqrt (2) ~.
\end{array}
\label{eq:expect1}
\end{equation}
{So,}  this is consistent with (\ref{eq:CHSH-Born1}) and (\ref{eq:CHSH-Born}) and therefore we don't have a contradiction.
The step from (\ref{eq:Bell-inequality1}) to (\ref{eq:Bell-inequality}) is allowed for expectations,
but is incorrect for {\em {conditional}}  expectations under different conditions. This is the cause of the earlier contradiction. 
Aside:  conditional expectations looks like a simple and innocent operation:
$ \EE[X+ Y \!\cc\! H_1] = \EE[X \!\cc\! H_2] + \EE[Y \!\cc\! H_3] $, is only true if $H_1=H_2=H_3$.
As an example consider the sum of dice $X$ and $Y$.
With $ \EE [X]=\EE[Y]=3.5$  it follows that $ \EE[X + Y]=7$.
However, $ \EE [X \plus Y  \cc X \eq Y] ~=~ \EE [X \plus Y  \cc  X \!\neq\! Y] ~=~ 7$.
Therefore, $\EE [X \plus Y]  ~\neq~  \EE [X \plus Y  \cc  X \eq Y]  + \EE [X \plus Y  \cc  X \!\neq\! Y]$.

We will explain this contradiction in a different way.
Define, for~each $\alpha$ and $\beta$, new random variables:
\begin{equation}
\begin{array}{lll}
\tilde{X}_\alpha &:= &X \mdot \IF{\{A = \alpha\}}  \\[0.5mm]
\tilde{Y}_\beta   &:= &Y\mdot \IF{\{B = \beta\}}.
\end{array}
\end{equation}
{It}  is easy to verify that:
\begin{equation}
\begin{array}{lll}
\EE[\tilde{X}_\alpha \tilde{Y}_\beta]
&=&\sum_{x,y} x\cdot y \cdot \PP[X\!=\!x, Y\!=\!y, A\!=\!\alpha ,B\!=\!\beta] 						\\[0.5mm]
&=&\sum_{x,y} x\cdot y \cdot \PP[X\!=\!x, Y\!=\!y\cc A\!=\!\alpha, B\!=\!\beta] \cdot \PP[A\!=\!\alpha, B\!=\!\beta]	\\[0.5mm]
&=&\EE[XY\cc A\!=\!\alpha, B\!=\!\beta] \cdot\PP[A\!=\!\alpha, B\!=\!\beta]					\\[0.5mm]
&=& \lh X_\alpha \tens Y_\beta \rh \cdot \PP[A\!=\!\alpha, B\!=\!\beta]
\end{array}
\label{eq:Bell-inequality-2}
\end{equation}
{If}  we take the four Tsirelson's angles  (\ref{angles}) for $\alpha$ and $\beta$
and we assume $\PP[A\!=\!\alpha, B\!=\!\beta]=\frac{1}{4}$ then we obtain
\begin{equation}
\begin{array}{lll}
& \big|\EE [\tilde{X}_0 \tilde{Y}_0] + \EE [\tilde{X}_1 \tilde{Y}_0] + \EE[\tilde{X}_1 \tilde{Y}_1] -\EE[\tilde{ X}_0 \tilde{Y}_1]\big| \\[1 mm]
=& \frac{1}{4}\big| \lh X_0 \tens Y_0 \rh + \lh X_1 \tens Y_0 \rh + \lh X_1\tens  Y_1 \rh -  \lh X_0 \tens Y_1 \rh \big| 			\\[1 mm]
=& \frac{1}{2}\sqrt 2\leq 2
\label{eq:Bell-inequality-3}
\end{array}
\end{equation}
{So,}  there is no contradiction with the four observables $\tilde{X}_0, \tilde{X}_1,\tilde{Y}_0,\tilde{Y}_1$.
But note that they are only different from $0$ if the settings are~correct.

Empirically, one has to estimate $\EE[\tilde{X}_\alpha \tilde{Y}_\beta]$,
which can be done by application of the law of large numbers:
\begin{equation}
\EE[\tilde{X}_\alpha \tilde{Y}_\beta]
~=~ \lim_{N \rightarrow \infty } \frac{1}{N} \sum_{n=0}^{N-1} X^{(n)} \mdot Y^{(n)}  \mdot \IF{\{A = \alpha, B = \beta\}} ~,
\end{equation}
where $X^{(n)}$ is an independent copy of $X$ and similarly $Y^{(n)}$ of $Y$.
Most likely, experimenters have computed:
\begin{equation}
\EE[XY \cc A=\alpha, B=\beta]
~=~ \lim_{N \rightarrow \infty } \frac{1}{M} \sum_{n=0}^{N-1} X^{(n)} \mdot Y^{(n)}  \mdot \IF{\{A = \alpha, B = \beta\}} ~,
\end{equation}
where
\begin{equation}
M:= \sum_{n=0}^{N-1}  \IF{\{A \!=\! \alpha, B \!=\! \beta\}}~,
\end{equation}
because it is more natural to take averages over the outcomes per setting.
This suggests that the experimental outcomes of $\lh X_\alpha \tens Y_\beta \rh$
correspond to conditional expectations, which explains the contradiction.
Below this model will be referred to as the {\em {2-observables model}}.

\section{The 2-Observables Model Extended with Hidden~Variables}
\label{sec:model-with-hidden-variables}

Following~\cite{Bell:1964}, we extend the proposed probability model with a hidden variable.
In order to accommodate this hidden variable,
the outcome space for the Bell experiment is extended with a random variable $\in \Lambda$,
which  is assumed here to be at most countable.
\begin{equation}
\Omega = \{-1,1\}^2 \times \{a_0,a_1\} \times \{b_0,b_1\}\times \Lambda~.
\label{eq:outcome-space-Lambda}
\end{equation}

Let random variable $L$ denote the hidden variable with $L(\omega)\!=\!\lambda$ for $\omega\!=\!(x,y,\alpha,\beta,\lambda)$.
The probability measure of this extended model must be consistent with Table~\ref{tab:conditional-probability-measure}
and hence must satisfy
\begin{equation}
\begin{array}{lll}
&  & \PP[X\!=\!x, Y\!=\!y \cc A\!=\!a, B\!=\!b]     							\\[1mm]
&=& \sum_{\lambda \in \Lambda} \PP[X\!=\!x, Y\!=\!y \cc A\!=\!a, B\!=\!b, L\!=\!\lambda]  \cdot
\PP[A\!=\!a, B\!=\!b, L\!=\!\lambda] ~.
\end{array}
\label{eq:probability-hidden-variables}
\end{equation}
{Critically,}  $L$ must be  {\em {independent}}  of the detector settings $A$ and $B$.
This assumption is known as {\em {statistical independence}}  or {\em {measurement independence}}  \cite{Hall:2015}:
\begin{equation}
\PP[A\!=\!a, B\!=\!b, L\!=\!\lambda] ~=~ \PP[A\!=\!a, B\!=\!b]  \cdot  \PP[L\!=\!\lambda]  ~.
\label{eq:measurement-independence}
\end{equation}
{This}  assumption is consistent with the idea that $\lambda$ is given before $A$ and $B$ are selected.
In this section we~prove:
\begin{enumerate}
\item This extended probability model is non-separable.
\item The non-separability of the model implies that it is non-deterministic or non-local (or~both).
\end{enumerate}
These properties are according to established~definitions.

\subsection{Statistical~Locality}
\label{subsec:statistical_locality}

With (\ref{eq:locality}) we observed that the outcome $X$ does not depend on the setting of detectors $A$ and $B$,
and likewise for outcome $Y$.
The property of  {\em {(statistical) locality}}  \cite{Hall:2015} is defined by
\begin{equation}
\begin{array}{lll}
\PP[X\!=\!x \cc A\!=\!\alpha, B\!=\!\beta, L\!=\!\lambda]	&=&	\PP[X\!=\!x \cc A\!=\!\alpha, L\!=\!\lambda]  \\[1mm]
\PP[Y\!=\!y\cc A\!=\!\alpha, B\!=\!\beta, L\!=\!\lambda]	&=&	\PP[Y\!=\!y \cc B\!=\!\beta, L\!=\!\lambda] ~.
\end{array}
\label{eq:statistically-local}
\end{equation}
{Note}  that this must hold for all $\alpha, \beta, \lambda$.
Paraphrasing~\cite{Hall:2015}:
``Thus, an~observer cannot distinguish, via any local measurement based on detector setting $A$,
whether a distant observer has carried out measurement using detector setting $B$ or $B'$),
even given knowledge of the underlying variable $\lambda$ .
This property, also known as {\em {parameter independence}}  and as  {\em {no-signalling}}  \cite{Brunner:2014},
is justified by the principle of relativity when the measurement subprocedures are carried out in spacelike separated regions~\cite{Bell:1964}.''
Statistical locality is not assumed a priori to be a property for the extended model including hidden~variable.

\subsection{Bell~Separability}
\label{subsec:bell-separability}

The probability model including a hidden variable, is {\em {separable}}  \cite{Bell:1964, Hall:2015}
if, for~all $x, y, \alpha, \beta$:
\begin{equation}
\begin{array}{lll}
p(x,y|\alpha,\beta)
&=&  \sum_{\lambda \in \Lambda}p_1(x|\alpha,\lambda) \cdot p_2(y|\beta,\lambda) \cdot \rho(\lambda)  \\[1mm]
p_1(x | \alpha, \lambda)  	&:=&  \PP[X \eq x \cc A \eq \alpha, L\!=\!\lambda]   \\[1mm]
p_2(y | \beta, \lambda) 	&:=&  \PP[Y \eq y \cc B \eq \beta, L\!=\!\lambda] ~,
\end{array}
\label{eq:separability}
\end{equation}
where $p(x,y|\alpha,\beta)$ is used as a shorthand for $\PP[X \eq x , Y\eq y\;|\;  A \eq \alpha, B \eq \beta] $,
and  $\rho(\lambda)$ is a shorthand for $\PP[L=\lambda]$.
The extended probability measure is repeated in Table~\ref{tab:space-abl},
including an assumed  factorization given hidden variable $\lambda$.
\begin{table}[htbp]
 \centering
 \renewcommand{\arraystretch}{1.2}         
     \begin{tabular}{cc|ccc}                     	\cline{1-4}
      		&	&   	$a$ 				&   	$a$ 			&	$A$  			\\ 
     		&	&   	$b$ 				&   	$b$ 			&	$B$  			\\ 
    $X$ & $Y$ 	&  	 				&  	$\lambda$ 	& 	$L$ 			\\ \cline{1-4}    
    1 &  1 	& $\frac12\sin^2(a \minus b)$	& $p(a,\lambda) 	\cdot  (1-q(b,\lambda))$ 	\\[0.5mm]  
   -1 &  1 	& $\frac12\cos^2(a \minus b)$	& $(1-p(a,\lambda)) 	\cdot  (1-q(b,\lambda))$ 	\\[0.5mm] 
    1 & -1 	& $\frac12\cos^2(a \minus b)$	& $p(a,\lambda) 	\cdot  q(b,\lambda)$ 		\\[0.5mm] 
   -1 & -1 	& $\frac12\sin^2(a \minus b)$	& $(1-p(a,\lambda)) 	\cdot  q(b,\lambda)$ 		\\[0.5mm]  \cline{1-4}
  \end{tabular}
 \caption{The conditional probability measure, given detector settings $a$ and $b$ (column $ab$),
 	      and an assumed factorization given hidden variable $\lambda$ (column $ab\lambda $).}
  \label{tab:space-abl}
\end{table}Separability, for~$\alpha = \beta$ would imply
\begin{equation}
\begin{array}{lll}
\sum_{\lambda \in \Lambda} ~ p_1(1|\alpha,\lambda) 		\cdot   p_2(1|\alpha,\lambda)\Big) \rho(\lambda) &=& 0 \\
\sum_{\lambda \in \Lambda} ~ (1-p_1(1 | \alpha, \lambda)) 	\cdot (1-p_2(1 | \alpha, \lambda))\rho({\lambda}) &=& 0 ~.
\end{array}
\label{sep1}
\end{equation}
{Hence,}  $1-p_1(1| \alpha, \lambda) -p_2(1| \alpha, \lambda)=0$
for all  $\lambda \!\in\! \Lambda$, with~$\rho(\lambda)\!>\!0$.
Substitution of $p_2(1| \alpha, \lambda)=1-p_1(1| \alpha, \lambda)$ in (\ref{sep1}) gives
\begin{equation}
\sum_{\lambda \in \Lambda} p_1(1|\alpha,\lambda)\cdot\Big(1 - p_1(1|\alpha,\lambda)\Big)\rho(\lambda) = 0 ~.
\label{sep3}
\end{equation}
{This}  constraint on $p_1$ can be rephrased as
\begin{equation}
p_1(x | \alpha, \lambda) 	\in \{ 0,1\} 	\hspace{2cm}
p_2(y | \beta, \lambda) 	\in \{ 0,1\}  ~.
\label{eq:p1p2-property}
\end{equation}
{This}  property relates to (non-)determinism, which is the topic of Section \ref{subsec:determinism}.

If Bell separability holds, we {derive:}  
\begin{equation}
\sum_{x,y}{x\cdot~y\cdot~p(x,y|\alpha,\beta)} ~=~
\sum_{\lambda \in \Lambda} \Big(\sum_{x}{x\cdot~p_1(x|\alpha,\lambda)} \cdot
\sum_{y}y\cdot~p_2(y|\beta,\lambda)\Big) \cdot \rho(\lambda)~,
\end{equation}  or, equivalently,
\begin{equation}
\EE[XY|A=\alpha, B=\beta] ~=~
\sum_{\lambda \in \Lambda}\EE[X|A=\alpha,L=\lambda] \cdot \EE[Y|B=\beta,L=\lambda] \cdot \rho(\lambda) ~.
\label{eq:E-seperability}
\end{equation}
{Next,}  by~choosing the elements $a_0,a_1,b_0,b_1$ equal to Tsirelson's angles (\ref{angles}) and we define
\begin{equation}
\begin{array}{llll}
Q_\lambda  &:=& \EE[X|A\!=\!a_0,L=\lambda] \\
R_\lambda  &:=& \EE[X|A\!=\!a_1,L=\lambda] \\
S_\lambda  &:=& \EE[Y|B\!=\!b_0,L=\lambda] \\
T_\lambda  &:=& \EE[Y|B\!=\!b_1,L=\lambda]~.
\end{array}
\end{equation}
{Note}  that $Q_\lambda,R_\lambda,T_\lambda,$ and $S_\lambda,$ are all in $[-1,1]$.
Their substitution in (\ref{eq:E-seperability}) results in.
\begin{equation}
\begin{array}{lll}
\EE[XY|A\!=\!a_0, B\!=\!b_0] 	&=& 	\sum_\lambda Q_\lambda S_\lambda ~\rho(\lambda) 	\\[1mm]
\EE[XY|A\!=\!a_0, B\!=\!b_1]	&=& 	\sum_\lambda Q_\lambda T_\lambda ~\rho(\lambda) 	\\[1mm]
\EE[XY|A\!=\!a_1, B\!=\!b_1] 	&=& 	\sum_\lambda R_\lambda S_\lambda ~\rho(\lambda)		\\[1mm]
\EE[XY|A\!=\!a_1, B\!=\!b_1]	&=& 	\sum_\lambda R_\lambda T_\lambda ~\rho(\lambda)~.
\end{array}
\end{equation}
{By}  (\ref{eq:expect1}) we have
\begin{equation}
\EE[XY|A\!=\!a_0, B\!=\!b_0]=\frac{1}{2}\sqrt2~,
\end{equation}
and similar for the others. So, on~the one hand
\begin{equation}
\begin{array}{lllll}
& \EE[XY|A\!=\!a_0, B\!=\!b_0] 	&+& 	\EE[XY|A\!=\!a_0, B\!=\!b_1]  			\\[1mm]
+& \EE[XY|A\!=\!a_1, B\!=\!b_1] 	&-& 	\EE[XY|A\!=\!a_0, B\!=\!b_1]) = 2\sqrt 2 ~,
\end{array}
\label{eq:CHSH2}
\end{equation}
and on the other hand the left-hand side of (\ref{eq:CHSH2}) equals:
\begin{equation}
\sum_\lambda \big(Q_\lambda S_\lambda+R_\lambda T_\lambda+R_\lambda S_\lambda-Q_\lambda T_\lambda\big) \cdot \rho(\lambda) ~.
\label{eq:CHSH3}
\end{equation}
{By}  the CHSH inequality (\ref{eq:CHSH-inequality})  for each  $\lambda \in \Lambda$:
\begin{equation}
|Q_\lambda S_\lambda+R_\lambda T_\lambda+R_\lambda S_\lambda-Q_\lambda T_\lambda|\leq 2 ~.
\end{equation}
{And}  therefore:
\begin{equation}
\sum_{\lambda \in \Lambda} \Big(Q_\lambda S_\lambda+R_\lambda T_\lambda+R_\lambda S_\lambda-Q_\lambda T_\lambda\Big) \cdot \rho(\lambda)\leq 2 ~.
\label{eq:CHSH4}
\end{equation}
{Equation}~(\ref{eq:CHSH2}) is in clear contradiction with (\ref{eq:CHSH4}).
This contradiction proves that the probability measure  is {\em {not}}  Bell separable.
This proof is a formalization of a sketch offered by~\cite{Brunner:2014}.
An alternative proof without the use of the CHSH-inequality is given in \mbox{Appendix \ref{alternative}.}
In summary we conclude that there does not exist a hidden variable that makes our probability measure separable.
Note that we did not assume any structure of the hidden variable, except~measurement independence
(\ref{eq:measurement-independence}).

\subsection{Determinism}
\label{subsec:determinism}

The model is called {\em {deterministic}}  if  for all $\lambda \in \Lambda$  \cite{Hall:2015}
\begin{equation}
\PP[X\!=\!x,Y\!=\!y \cc A\!=\!a, B\!=\!b, L\!=\!\lambda] ~\in~ \{0,1\} ~.
\label{eq:determinism}
\end{equation}
{For}  a model that is (statistically) local (\ref{eq:statistically-local}),
Bell separability (\ref{eq:separability}) is equivalent to \mbox{determinism:}
\begin{equation}
\cal S \land \cal L ~\Leftrightarrow~  \cal D \land \cal L ~,
\label{eq:theorem-determinism}
\end{equation}
where  $\cal S, \cal L$, and~$\cal D$ denote the properties separable, statistically local, and~deterministic, respectively.
The immediate implication is
\begin{equation}
\neg  \cal S~\Rightarrow~  \neg \cal D \lor \neg \cal L ~.
\label{eq:determinism-implication}
\end{equation}
In words, the~proposed probability model is not separable, and~therefore it is non-deterministic or non-local (or both).

The proof of (\ref{eq:theorem-determinism}) is presented in two parts.
First we focus on $\cal S \land \cal L ~\Rightarrow~  \cal D$, which makes use of the following lemma.
\begin{equation}
\begin{array}{llll}
& p_1(x | \alpha, \beta, \lambda) \in \{ 0,1\} 	~\land~	p_2(y | \alpha, \beta, \lambda)  \in \{ 0,1\}  \\[1mm]
\Rightarrow	& p(x ,y| \alpha, \beta, \lambda)  ~=~ p_1(x | \alpha, \beta, \lambda) \cdot p_2(y | \alpha, \beta, \lambda)
~\in \{ 0,1\} ~,
\end{array}
\label{eq:determinism-factorize}
\end{equation}
where $p$, $p_1$, and~$p_2$ are obvious generalizations of those in (\ref{eq:separability}).
From $p_1()$ we conclude that for only one $x$ we have $p_1(x| \alpha, \beta, \lambda) \neq 0$ and similarly for $p_2()$.
Hence, for~only one pair $x,y$ it holds that $p(x ,y| \alpha, \beta, \lambda) \neq 0$. For~that pair the probability must be~1.

Given this lemma, we prove $\cal S \land\cal  L ~\Rightarrow~  \cal D~$ by
\begin{equation}
\begin{array}{lll}
& \cal S \land \cal L 	 														\\[1mm]
\Rightarrow 	&  \{ \textrm{Property} ~(\ref{eq:p1p2-property}) \}					 				\\[1mm]
& p_1(x | \alpha, \lambda) \in \{ 0,1\}  ~\land~ p_2(y | \beta, \lambda) \in \{ 0,1\} 	~\land~ \cal L 	\\[1mm]
\Rightarrow 	&  \{ \textrm{Definition} ~{\cal L}, ~(\ref{eq:statistically-local}) \}					 		\\[1mm]
& p_1(x | \alpha, \beta, \lambda) \in \{ 0,1\}  ~\land~ p_2(y |\alpha,  \beta, \lambda) \in \{ 0,1\} 	\\[1mm]
\Rightarrow 	&   \{ \textrm{Property} ~(\ref{eq:determinism-factorize}) \}				 				\\[1mm]
& p(x,y | \alpha, \beta, \lambda) \in \{ 0,1\}   										\\[1mm]
= 			&  \{ \textrm{Definition} ~{\cal D}, ~(\ref{eq:determinism}) \}					 		\\[1mm]
& \cal D   ~.
\end{array}
\end{equation}

The  second part of the proof makes use of a similar lemma.
\begin{equation}
\begin{array}{lll}
& p(x ,y| \alpha, \beta, \lambda)  ~\in~ \{ 0,1\} 									\\[1mm]
\Rightarrow	& p_1(x | \alpha, \beta, \lambda) \in \{ 0,1\} ~\land~ p_2(y | \alpha, \beta, \lambda) \in \{ 0,1\}  	\\[1mm]
& ~\land~ p(x ,y| \alpha, \beta, \lambda) ~=~ p_1(x | \alpha, \beta, \lambda) \cdot p_2(y | \alpha, \beta, \lambda) ~,
\end{array}
\label{eq:determinism-factorize2}
\end{equation}
because there can be only one $x,y$ for which $p(x ,y| \alpha, \beta, \lambda) =1$.

Given this lemma, we prove $\cal D \land \cal L ~\Rightarrow~  \cal S~$ by
\begin{equation}
\begin{array}{lll}
& p(x ,y| \alpha, \beta) 			\hspace{1cm} \textrm{assume} ~\cal D \land \cal L 			\\[1mm]
= 	&  \{ \textrm{Probability theory and measurement independence}  ~(\ref{eq:measurement-independence}) \}	\\[1mm]
& \sum_{\lambda \in \Lambda} p(x,y | \alpha, \beta,\lambda) \cdot \rho(\lambda)
\hspace{1cm} \textrm{assume} ~\cal D \land \cal L  	\\[1mm]
= 	&   \{ \textrm{Property} ~(\ref{eq:determinism-factorize2}) \textrm{, given}~\cal D \}			\\[1mm]
& \sum_{\lambda \in \Lambda} p_1(x | \alpha, \beta,\lambda) \cdot p_2(y | \alpha, \beta,\lambda)
\cdot \rho(\lambda) \hspace{1cm}  \textrm{assume} ~ \cal L  			\\[1mm]
= 	&  \{ \textrm{Definition} ~{\cal L}, ~(\ref{eq:statistically-local}) \}					 			\\[1mm]
& \sum_{\lambda \in \Lambda} p_1(x | \alpha, \lambda) \cdot p_2(y | \beta,\lambda)
\cdot \rho(\lambda)  ~. \hspace{1cm}
\end{array}
\end{equation}
{That}  is, given $\cal D \land \cal L$, the~probability model is Bell-separable (\ref{eq:separability}).

\section{Conclusions}
\label{sec:conclusion}

This paper proposes a probability model for the Bell experiment.
The outcome space is defined as the cartesian product space of random variables $A$ and $B$
(detector settings) and $X$ and $Y$, the~corresponding measurement outcomes.
It has only two simultaneously observable detector settings per measurement
and therefore does not need the assumption of realism.
The probability measure $p(x, y, a, b)$ is derived from quantum mechanics and described by
Tables~\ref{tab:conditional-probability-measure} and \ref{tab:CHSH-probability-measure}.

From this probability measure it follows that the
{\em {conditional}}  expectation\linebreak   $ \EE[X \mdot Y \Cc A \!=\! \alpha, B \!=\! \beta ]$
corresponds to Born expectation $\lh X_\alpha \tens Y_\beta \rh$.
The Bell contradiction arises from the summation of conditional expectations
with different conditions (i.e., different detector settings).
The 2-observables model (Section \ref{suitable model}) does not show such a contradiction.
The experimental results by Aspect et al.~\cite{Aspect:1982} are
``in excellent agreement with the quantum mechanical predictions''
and therefore also with the proposed model.

In Section~\ref{sec:model-with-hidden-variables} we extend the probability model
with a hidden variable.
This extended probability model is proved to be non-separable in the Bell sense. ({{The}  
proof in Section~\ref{subsec:bell-separability} makes use of the CHSH inequality.
The alternative proof in Appendix~\ref{alternative} does not.)}
This non-separability of the probability measure supports Bell's~\cite{Bell:2004} conclusion
``If [a hidden-variable theory] is local it will not agree with quantum mechanics,
and if it agrees with quantum mechanics it will not be local''.
Furthermore, non-separability implies that the model is non-deterministic or non-local (or both).

\acknowledgments{
The~authors thank Richard Gill (Leiden University),
Andrea Fiore, Servaas Kokkelmans, and~Henk Nijmeijer (all Technical University Eindhoven),
and  Sven Jandura, (University of Strasbourg)
for providing helpful feedback on an earlier version of this paper.
The feedback of the anonymous reviewers helped us
to greatly improve the presentation of the material in Section~\ref{sec:model-with-hidden-variables}.}
\appendix

\section{Alternative Proof for~Non-Separability}
\label{alternative}
The argument below focuses on the specific measurement outcome $(X=+1, Y=-1)$.
We introduce new shorthands $p(\alpha, \lambda )$ and $q(\beta, \lambda)$ based on (\ref{eq:separability}):
\begin{equation}
\begin{array}{ll}
p(\alpha, \lambda) &= p_1(+1|\alpha,\lambda) \\
q(\beta, \lambda) &=p_2(-1|\beta,\lambda)  ~.
\end{array}
\label{eq:bell-separable-pq-def}
\end{equation}
{Table}~\ref{tab:conditional-probability-measure}
implies that the outcomes of single-sided experiments must satisfy
\begin{equation}
\int_\Lambda p(\alpha, \lambda)  \rd \rho(\lambda)=\frac12 ~~~~ \text{and} ~~~~
\int_\Lambda q(\beta, \lambda)  \rd  \rho(\lambda)=\frac12  ~.
\label{eq:bell-separable-pz-qz}
\end{equation}
{With}  this notation Table~\ref{tab:space-abl} is adapted into Table~\ref{tab:space-abl2}.
\begin{table}[htbp]
 \centering
 \renewcommand{\arraystretch}{1.2}         
     \begin{tabular}{cc|ccc}                     	\cline{1-4}
      		&	&   	$a$ 				&   	$a$ 			&	$A$  			\\ 
     		&	&   	$b$ 				&   	$b$ 			&	$B$  			\\ 
    $X$ & $Y$ 	&  	 				&  	$\lambda$ 	& 	$L$ 			\\ \cline{1-4}    
    1 &  1 	& $\frac12\sin^2(a \minus b)$	& $p(a,\lambda) 	\cdot  (1-q(b,\lambda))$ 	\\[0.5mm]  
   -1 &  1 	& $\frac12\cos^2(a \minus b)$	& $(1-p(a,\lambda)) 	\cdot (1-q(b,\lambda))$ 	\\[0.5mm] 
    1 & -1 	& $\frac12\cos^2(a \minus b)$	& $p(a,\lambda) 	\cdot q(b,\lambda)$ 		\\[0.5mm] 
   -1 & -1 	& $\frac12\sin^2(a \minus b)$	& $(1-p(a,\lambda)) 	\cdot q(b,\lambda)$ 		\\[0.5mm]  \cline{1-4}
  \end{tabular}
 \caption{The conditional probability measure, given detector settings $a$ and $b$ (column $ab$),
 	      and an assumed factorization given hidden variable $\lambda$ (column $ab\lambda $).}
  \label{tab:space-abl2}
\end{table}

For specific outcome $(+1, -1)$ Equation~(\ref{eq:Pxy-b}) states that
$\PP[+1,-1 \Cc \alpha, \beta]=\frac12\cos^2(\alpha \minus \beta)$, hence
Bell-separability implies:
\begin{equation}
\int_\Lambda p(\alpha, \lambda) \cdot q( \beta, \lambda) \cdot \rd \rho(\lambda)=\frac12  \cos^2 (\alpha \minus \beta)  ~.
\label{eq:bell-separable-pq}
\end{equation}
{Below}  we proceed with the assumption that such $p$ and $q$ do exist and derive a contradiction in five steps.
(Here we don't assume $\Lambda$ to be countable, so we use integration instead of~summation.)

{\bf {Step 1:}} 
rewrite the above conditions in terms of $\tilde{p}, \tilde{q}$,
using $\tilde{p} = 2p \!-\!1$, $\tilde{q} = 2q\! -\!1$, and~$\cos(2(\alpha  \minus \beta)) = 2 \cos^2 (\alpha  \minus \beta) \!-\!1$:
\begin{equation}
\begin{array}{lll}
\tilde{p},\tilde{q} : [0,\pi]\times \Lambda \to[-1,1]~,           		         							& & (i)	\\[1.5mm]
\int_\Lambda \tilde{p}(\alpha , \lambda )  \rd \rho(\lambda) =0,~~~~
\int_\Lambda \tilde{q}(\alpha , \lambda )  \rd \rho(\lambda) =0~,									& & (ii)	\\[1.5mm]
\int_\Lambda \tilde{p}(\alpha , \lambda )\tilde{q}(\beta, \lambda)  \rd \rho(\lambda) =\cos\big(2(\alpha  \minus \beta)\big) ~.	& & (iii)
\label{eq:bell-separable-conditions}
\end{array}
\end{equation}

{\bf {Step 2:}}   demonstrate that  $\tilde{p}=\tilde{q}~$ and that  $\tilde{p}(\alpha , \lambda) \in \{-1,1\}$.
These properties follow from taking $a \eq b$ and inspecting of the following integrals:
\begin{equation}
\int_\Lambda(1+\tilde{p}(\alpha, \lambda ) )(1-\tilde{q}(\alpha , \lambda ) )\rd\rho(\lambda) =0,~~~~
\int_\Lambda(1+\tilde{q}(\alpha, \lambda ) )(1-\tilde{p}(\alpha , \lambda ) )\rd\rho(\lambda) =0 ~.
\end{equation}
{The}  factors in the two integrands are all non-negative and therefore,
\begin{equation}
1 + \tilde{p}(\alpha, \lambda ) - \tilde{q}(\alpha , \lambda ) - \tilde{p}(\alpha , \lambda ) \tilde{q}(\alpha , \lambda )=0,~~~~
(\alpha, \lambda ) - \tilde{p}(\alpha , \lambda ) - \tilde{p}(\alpha , \lambda ) \tilde{q}(\alpha , \lambda )=0 ~,
\end{equation}
which leads to $\tilde{p}(\alpha, \lambda )=\tilde{q}(\alpha , \lambda ) ~.$
As a result of this equality $~\int_\Lambda (1-\tilde{p}^2(\alpha , \lambda ) )\rd\rho(\lambda) =0~$,
and $\tilde{p}(\alpha , \lambda )$ can necessarily only assume values in $\{-1,1\}$. ({{Assuming}  
that such $\tilde{p}$  and $\tilde{q}$ exist,  this constraint implies deterministic measurement outcomes.})

{\bf {Step 3:}}  rephrase the conditions, assuming the orthonormal bases
\begin{equation}
\big\{\, \alpha		\mapsto	\frac{e^{2\ri k \alpha}}{\sqrt\pi}\big\}_{k\in\ZZ} ~~~~~~~~~~~~
\big\{(\alpha, \beta)	\mapsto	\frac{e^{2\ri k \alpha}e^{2\ri \ell \beta}}{ \pi}\big\}_{(k,\ell)\in\ZZ^2} ~.
\end{equation}
{Next,}  write $\tilde{p}$ in terms of this Fourier basis
\begin{equation}
\tilde{p}(\alpha, \lambda )=\sum_{k\in\ZZ} \,  c_k(\lambda)\frac{e^{2\ri k \alpha}}{\sqrt\pi},~~~\text{where }~~~c_k(\lambda)=\int_0^\pi\tilde{p}(\alpha, \lambda )\frac{e^{-2\ri k \alpha}}{\sqrt\pi}\rd \alpha ~,
\end{equation}
and use
\begin{equation}
\cos\big(2(\alpha \minus \beta)\big)=\frac\pi2\frac{e^{2 \ri \alpha}e^{-2 \ri \beta}}{\pi}+
\frac\pi2\frac{e^{-2 \ri \alpha}e^{2 \ri \beta}}{\pi}
\end{equation}
to rewrite (\ref{eq:bell-separable-conditions}) {$(ii)$}  and {$(iii)$}  as
\begin{equation}
\begin{array}{lll}
\int_\Lambda \tilde{p}(\alpha, \lambda )  \rd \rho(\lambda)
=\sum_{k\in\ZZ}\Big(\int_\Lambda c_k(\lambda) \rd\rho(\lambda) \Big)\frac{e^{2\ri k \alpha} }{\sqrt\pi}=0,    & & (ii) \\[2mm]
\sum_{(k,\ell)\in\ZZ^2}\Big(\int_\Lambda c_k(\lambda)c_l (\lambda)\rd\rho(\lambda) \Big)\frac{e^{2\ri k \alpha}e^{2\ri \ell \beta}}{ \pi}~=~
\frac\pi2\frac{e^{2 \ri \alpha}e^{-2 \ri \beta}}{\pi} +	\frac\pi2\frac{e^{-2 \ri \alpha}e^{2 \ri \beta}}{\pi}. & & (iii)
\end{array}
\end{equation}

{\bf {Step 4:}}  derive the conditions on $k, \lambda \mapsto c_k(\lambda)$,
using the orthogonality of the Fourier expansions:
\begin{equation}
\begin{array}{rll}
c_0(\lambda)\in\RR,~~~\forall k\in\ZZ: 		& \int_\Lambda c_k(\lambda)\rd\rho(\lambda)  				& = 0. 	 	\\[2mm]
& \int_\Lambda c_1(\lambda)c_{-1}(\lambda)\rd\rho(\lambda) 	& = \frac\pi2 	\\[2mm]
\forall (k,\ell)\in\ZZ^2\backslash\{(1,-1),(-1,1)\}:	& \int_\Lambda c_k(\lambda)c_l(\lambda)\rd\rho(\lambda)  	& = 0		\\[2mm]
\forall k \in\ZZ\backslash\{-1,1\}:                       	& \int_\Lambda c_k(\lambda)c_{-k}(\lambda)\rd\rho(\lambda)  	& = 0 		~.
\end{array}
\end{equation}
{The}  final condition is a subtly chosen subset of the preceding one.
Also note that all $c_k(\lambda),~|k|\neq1$, have to vanish
and that $~c_k(\lambda)c_{-k}(\lambda) = |c_k(\lambda)|^2~$, since $~c_k(\lambda)=\overline{c_{-k}(\lambda)}$.
Summarizing for the remaining $|k|=1$,
the following conditions are required to solve (\ref{eq:bell-separable-conditions}):
\begin{equation}
\int_\Lambda    c_{\pm1}(\lambda) \rd\rho(\lambda) =0, ~~~~\int_\Lambda  ( c_{\pm1}(\lambda))^2\rd\rho(\lambda) =0,~~~~\int_\Lambda  |c_{\pm1}(\lambda)|^2\rd\rho(\lambda) =\frac\pi2 ~.
\label{eq:ak_condition}
\end{equation}

{\bf {Step 5:}}  expose the contradiction.
The analysis above leads to
\begin{equation}
\tilde{p}(\alpha, \lambda )=\dfrac{1}{\sqrt{\pi}}\big(c_1(\lambda)e^{2 \ri \alpha}+c_{-1}(\lambda)e^{-2 \ri \alpha}\big)~
= ~  \dfrac{ 2 | c_1(\lambda)| }{\sqrt{\pi} } \cos \big(2 \alpha + \alpha(\lambda) \big) ~,
\end{equation}
where $\alpha (\lambda)$ is given by $c_1 =  | c_1(\lambda) | e^{- i \alpha (\lambda)} $.
The values of $c_{\pm1}(\lambda)$ depend on the choice for $\Lambda$. For~example,
if $\Lambda=[0,1]\subset\RR$ then condition (\ref{eq:ak_condition}) is met by
$c_{\pm1}(\lambda)=\sqrt{\frac\pi2}e^{\pm2\pi\ri \lambda}$.
Note that $\tilde{p}(\alpha, \lambda )$ assumes many more values than just $\pm1$ as was established in Step 2.
With this contradiction in the derived properties of $\tilde{p}(\alpha, \lambda )$ we conclude that
$p(x, y, \alpha, \beta)$ is {\em not} Bell~separable.


\bibliographystyle{ieeetr}
\bibliography{Bell-probability-model} 

\end{document}